\documentclass[12pt]{article}
\usepackage{multirow}
\usepackage{amsfonts}
\usepackage{amsmath}
\usepackage{graphicx}

\newcommand{\beq}{\begin{equation}}
\newcommand{\eeq}{\end{equation}}
\newcommand{\bea}{\begin{eqnarray}}
\newcommand{\eea}{\end{eqnarray}}

\newcommand{\Lab}{\rm{L}^*\rm{a}^*\rm{b}^*}

\begin{document}
\noindent {\Large Fractal Dimensions in Perceptual Color Space: A Comparison
Study Using Jackson Pollock's Art}
\vskip .3cm
\noindent {J.\ R.\ Mureika \\
{{\it Department of Physics, Loyola Marymount University}
\\ {\it 1 LMU Drive, Los Angeles, California~~90045-8227} \\
\footnotesize Email: jmureika@lmu.edu}}
\vskip .3cm
\vskip .5cm
\noindent Abstract\\
{\footnotesize
The fractal dimensions of color-specific paint patterns in various Jackson 
Pollock paintings are calculated using a filtering process which models 
perceptual response to color differences ($\Lab$ color space).  The 
advantage of the $\Lab$ space filtering method over traditional RGB spaces is 
that the former is a perceptually-uniform (metric) space, leading to a more 
consistent definition of ``perceptually different'' colors.  It is determined 
that the RGB filtering method underestimates the perceived fractal dimension
of lighter colored patterns but not of darker ones, if the same selection 
criteria is applied to each.  Implications of the findings to 
Fechner's 'Principle of the Aesthetic Middle' and Berlyne's work on
perception of complexity are discussed.

}

\vskip 1cm

{\bf 
The use of fractal analysis to explain aesthetic properties of art is
becoming a subject of great interdisciplinary interest to physicists,
psychologists, and art theorists.  Previous studies have addressed the classification
of abstract expressionist art by the fractal dimension of the pigment
patterns on the canvas as a method of artist authentication.  Moreover,
it has been proposed that the fractal structure of the pigment patterns
is somehow connected to the aesthetic ``value'' of the painting.  
The patterns in question have traditionally been selected using filtering
algorithms of RGB primaries, a perceptually non-uniform color space in
which ``distances'' between perceptually just-differentiable colors is
not the same for lighter and darker hues.  Although RGB-based analyses
have had success in devising categorization schemes for abstract paintings
(see the cited literature), the use of this color space
limits analysss which seek to cross-compare the fractal dimension of 
different color patterns from a {\it perceptual} stance.  
The following report summarizes the results of a fractal analysis performed 
on several paintings by the 
renowned artist Jackson Pollock, this time in a perceptually-uniform color 
space which more closely replicates how the visual cortex would identify 
and differentiate individual colors.   The data provides better insight into 
the fractal 
dimension and aesthetic nature of specific light and dark pigment patterns, 
and posits that the artist may have primarily used darked colors to engage 
the viewer.
}

\section{Fractals in Abstract Expressionist Art}
Fractals are implicitly tied to the notions of chaos and irregularity
\cite{mandel1,falconer,barnsley}, and over the past 15 years have been
increasingly associated with human perception issues.  
The problem of structure identification and discrimination in music,
art, and visual processing has benefitted greatly from this cross-disciplinary
endeavor.  For example, the authors of \cite{schmuckler1,schmuckler2} 
pose the question of whether or not
humans are ``attuned'' to the perception of fractal-like
optical and auditory stimuli.  Similarly, the results reported in \cite{gliden}
show that the quantitative accuracy of human memory possesses a fractal-like 
signature which can be measured in task repitition.  
Specifically, when subjects were asked to perform tasks such as 
repeatedly drawing lines of specific lengths or shapes, the statistical
variations in the lengths have been shown to be not purely random noise, but
fractally ordered ``$1/f$'' noise.  

Recently, the use of fractal dimension analysis techniques
for the study of paintings has become of interest \cite{taylor1,taylor2,
taylor3,taylor4,jrmleo,jrmpre}, which in the case of works by Jackson Pollock suggest
that the fractal dimension of the paint patterns cluster suspiciously around the
value $D_F \sim 1.7$.  In Reference~\cite{jrmleo,jrmpre}, the analysis is extended
to paintings by different artists and addresses the full multifractal
spectrum of the patterns. Furthermore, to overcome the problem of proper
color choice (the focus of discussion in this paper), 
the notion of a {\it visual fractal} was introduced \cite{jrmleo}. 
Instead of direct observation of colors, the focus instead shifted to
{\it edge structures}.  This is effectively an analysis of luminance gradients
within the image, and not directly on the RGB color field distribution.

Implicitly related to this
topic, the authors of \cite{gerry1}
discuss the perceptibility of hierarchical structures in abstract or
non-representational constructs.  In fact, rapid object
recognition and categorization via boundary isolation versus ``blob''
identification is a subject of growing scientific interest (see
\cite{schyns2} and related references therein).  Similarly,
the degree of complexity present in a scene is largely believe to be
critical in maintaining the interest of an observer \cite{berlyne1,berlyne2}. 
The fractal dimension is a natural measure of such complexity.

The predominant question remains: ``where is the fractal''?  Does one 
calculate this statistic based on a pattern of a specific color?  If so, 
how is this color selected and specified?  A simple choice would
be to pick the most abundant values of red, green, and blue (hereafter RGB)
primaries and digitally deconstruct the image to remove the appropriate
matching pieces.  Patterns which match this selection criteria can be called
``physical colors'', since the RGB primaries define the image as it appears
(on the canvas).

However, the human visual processing system has evolved in such a way that the
actual physical world is  not always what is perceived by the brain.
There is a long-standing argument addressing the questions of how we 
process scenes, what elements are important to a visual field, and so forth.
As previously mentioned, the analysis in References~\cite{jrmleo,jrmpre} studies the
edge structure of paintings, based on the notion that we perceive contrast
changes separately (or independently) from individual colors.  

%In fact, a complete understanding of the nature of color perception itself
%is still lacking.
Similarly, perceived differences between colors themselves are non-trivial to
quantify.  In fact, use of RGB primaries for {\it perceptual} image analysis 
is flawed because the color space in question is not perceptually uniform.
In this paper, previously-reported fractal dimensions for various paintings
by Jackson Pollock are re-computed using what will be termed {\it perceptual
color selection}, as opposed to {\it physical color selection}.  The latter
uses the simple RGB primaries, while the former involves computations in 
the CIE-$\Lab$ color space.

The following paper will analyze six paintings by Jackson Pollock by 
determining the fractal dimension of specific patterns 
formed in the $\Lab$ color space.  This data will be compared to the 
fractal dimensions of the same color patterns in the usual RGB color space,
and thus the results can be understood to represent the perceptual distinctions
of colors on the canvas.

\section{The Basics of Perception}
\label{colorvision}
Before attacking the problem of detecting visual fractals, a brief primer
on color vision and perception is in order.  In fact, it was physicists
who had the first major say in the foundations of this science, known
in the literature as ``psychophysics''.

In the early 1800s, the {\it Trichromacy Theory}
of vision was postulated by Thomas Young, and was later expanded upon 
by Helmholtz and Maxwell (later dubbed the Young-Helmholtz Theory, much to
the dismay of Maxwell) \cite{nassau}.  The assertion was that color vision is 
the result of simultaneous stimulation of three photoreceptors in the eye, 
based on the RGB primary breakdown.  Physiological confirmation of this 
hypothesis did not come until the 1960s, when three distinct cellular 
receptors in the eye (cones) were discovered to have peak sensitivities to
light of $\lambda = 440~$nm (blue), 540~nm (green), and 580~nm (actually
more yellow than red).

Meanwhile, the late 1800s saw the emergence of Karl Ewald Konstantin
Hering's {\it Opponent Theory} of vision \cite{nassau}.  Instead of a
trichromatic basis for vision, Hering proposed that the perception of
colors was derived from the contrasting of opposite color/intensity pairs:
red-green, yellow-blue, and light-dark.  Again, experimental physiological
evidence for such a mechanism was revealed in the 1950s.  In this case,
two chromatic signals and a third achromatic one were detected in the
optical nerve under various stimulation experiments.  

Note that unlike the
Trichromacy Theory, the Opponent Theory allows for object recognition based
on luminosity or hue gradients alone, and hence no explicit color 
information is required.  So, while the raw color stimuli may be perceived,
it may not be this information which is transmitted to the visual cortex
for eventual processing. 

Most modern theories of color perception tend to constitute a mixture of
the two aforementioned postulates in some fashion.  This, of course, leads
to the immediate question: is there a preferential order for object and
color detection?  Is one a primary mechanism, and the other secondary?  Or,
are they mutually independent processes which serve to provide diverse
information about the scene considered?  There is still no clear answer
to these musings,  although much work has been devoted to such studies
(see texts such as {\it e.g.} \cite{hcv} and references therein for 
further reading).

\section{CIE Color Systems}
\label{cie}
The {\it Commission Internationale de l'Eclairage}, or CIE as it is more often
known, was formed in an attempt to address and standardize the myriad aspects
of color definition, reproduction, and perception via a rigorous set of
mathematical standards and transformations.  Since actual color perception
can vary depending on the external conditions (ambient lighting) and
internal conditions of the observer (neurophysiology of vision mechanism),
a set of ``invariant'' standards is useful in describing ideal conditions
under which observations and comparisons can be made.

In order to establish consistent external lighting variables, the CIE
defined the Standard Illuminants to be those conditions which represent
the complete spectral power distribution of a particular state.  
The most widely-used of these standards are the D-illuminants, which 
characterize the conditions of ``average daylight''.  In the present work,
all CIE conversions will reference the D65 Illuminant, which corresponds to
standard average daylight with a spectral temperature of 6500~K 
\cite{nassau,graphics}.  Note that the D-Illuminants standards cannot be
reproduced by any known physical source of light.  Conversely, the
earlier A, B, and C-Illuminants were based on the spectral power distributions
of (filtered) incandescent tungsten light (2854~K) \cite{nassau}.  This
mild lack of chromatic reproducibility is an inherent problem with digital 
analyses of images, however with a 24-bit color system it is doubtful that
it constitutes a large concern.

It should be noted that CIE color systems are primarily designed for
industrial (textile) color-matching and color gamut consistency in color
displays.  While many of their intricacies are based on human perception
principles, they are not meant to fully represent the neural processes which
occur in vision.  For the purposes of this manuscript, however, they are certainly
a good first-pass approach at the problem.

\section{Filtering Visual Fractals}
To date, the color-filter process has relied on the fact that the target
colors are the mixture of RGB triplets.  Such a color basis is certainly
not unreasonable, and in fact forms a large base of the tristimulus theory
of color vision.  However, further inspection of color theory reveals
that the three-dimensional RGB space is {\it not perceptually uniform}.
That is, two colors which are a fixed distance $\beta_{RGB}$ away from a
base stimulus may not be equally different from a perceptual stance.

\subsection{Alternate Color Representations}
Furthermore, the RGB specification is deficient in the sense that,
as an additive color scheme, it cannot reproduce all observed colors.
In 1931, the CIE set out to formulate
an accurate color space.  Known as the CIE XYZ space, 
these {\it tristimulus} primaries themselves are not visible in the same
sense as R, G, and B, but are rather an ``imaginary'' basis introduced
to allow for reproduction of all observable colors.
Specific colors $C(\lambda)$ are matched by combining appropriate amounts of
red, green, and blue primaries (denoted r, g, and b).  However, in many
cases, it was noted that perfect matches
could not be made in such a fashion.  Instead, one could match combinations
of two of the three primaries with a suitable combination of the target
color and the third primary.  Arithmetically, this implies

\beq
C(\lambda) + r{\bf R} = b{\bf B} + g{\bf G} \\
\eeq
and so the target $C(\lambda)$ is formed by a negative contribution from
one of the primaries.  The CIE XYZ system thus reproduces the entire spectrum
of observable colors.

For a standard D65 Illuminant observer, the transformation is a simple
linear one of the form
\bea
\left( \begin{array}{c} X \\ Y \\ Z \end{array} \right) & = &
\left( \begin{array}{ccc}
0.412424& 0.212656 &0.0193324\\
0.357579& 0.715158 &0.119193\\
0.180464& 0.0721856 &0.950444
\end{array} \right) \;
\left( \begin{array}{c} R \\ G \\ B \end{array} \right)
\label{rgb2xyz}
\eea
with the inverse transform yielding negative coefficients, as indicated
above.  The exact form of the matrix in Equation~\ref{rgb2xyz} is somewhat
dependent on the color gamut and standard white being used for display
purposes.  In the case
of this paper, the matrix values are for the sRGB color scheme
(for ``standard RGB''), and will primarily be adopted for the analysis herein.
However, comparison with other transformation schemes will be discussed.

Unfortunately, while the XYZ space is more physically realistic in terms
of color reproducibility, it is still not perceptually uniform.  The CIE
addressed these issues, and offered several solutions as recently as 1976.

\subsection{CIE-$\Lab$ Space: Perceptual Uniformity}
\label{cielab}
A truly perceptually-uniform space, the CIE-$\Lab$ color space is a non-linear
transformation of the XYZ space:

\bea
\rm{L}^* & = & 116\; f(Y/Y_0) - 116  \nonumber \\
\rm{a}^* & = & 500 \left[f(X/X_0) - f(Y/Y_0)\right] \\
\rm{b}^* & = & 200 \left[f(Y/Y_0) - f(Z/Z_0)\right] \nonumber 
\label{labeq}
\eea
where $f(X/X_0) = (X/X_0)^\frac{1}{3}$ if $(X/X_0) > 0.008856$, and
$f(X/X_0) = 7.787\:(X/X_0)+16/116$ otherwise \cite{nassau}.  Here, the
values $(X_0,Y_0,Z_0)=(0.3127,0.3290,0.3583)$ are 
the standard (white) tristimulus values for a $2^\circ$ observer in the D65
illuminant (in general, one can make the approximation
$X_0=Y_0=Z_0 = 1/3$).
The coordinate $L^*$ represents the perceived luminosity, and covers the
range of luminance scales (0 being black, 100 being white).  The remaining
coordinates $a^*$ and $b^*$ are the relative red-green and blue-yellow
content, analogous to Hering's Color Opponent theory and more realistic
ocular color detection processes \cite{nassau}.

The perceptual color difference is then the Euclidean distance in $\Lab$ space,
\beq
\beta_{\Lab} = \sqrt{(\Delta L^*)^2 + (\Delta a^*)^2 + (\Delta b^*)^2}
\label{labdiff}
\eeq

One immediately notes from the form of Equation~\ref{labeq} that the structures
of the RGB and $\Lab$ color spaces are quite different.  This suggests that
the relative structures obtained by color-filter processes are largely
dependent on the color-matching system at hand.  Specifically, one might
expect that the patterns selected by RGB filtering criteria do not conform
to those of a $\Lab$ filter.  That is, the {\it physical} distribution of
like colors may not correspond to the {\it perceived} distribution of colors.
If the structures are sufficiently different, then this can weaken arguments
which suggest that patterns of specific fractal dimension are pleasing to
observers.

%The measured $D_q$ spectra for both light and dark colors can be effectively
%reproduced under this transformation, except that the two cannot have the
%same cutoff radius.  While this seems somewhat intuitive, the psychophysical
%implications could run deeper.  In RGB space, the measured dimensions were
%cited as $D_0 = 1.77, 1.34$ for the black and yellow patterns, 
%with $D_{\infty} =1.61, 1.14$ respectively. 

%However, two different cut-offs are required in $\Lab$.  The yellow $D_q$
%spectrum is roughly replicated for $\beta_{\Lab} \sim 7$, but using the
%same cut-off, the black pattern yields $D_0 = 1.67$, $D_{\infty} = 1.42$.
%Equivalently, $\beta_{\Lab} \sim 15$ gives the black RGB $D_q$ spectrum, 
%which for the yellow pigment now becomes $D_0 = 1.58$, $D_{\infty} = 1.31$.

The difference in measured spectra may indeed by a visual effect, if the
eye functions on a similar uniform ``cut-off'' level for like-color
discrimination.  However, the actual color information of the system
may not be the most important contributor to first order visual processing
systems.

\section{Analysis and Results}
The images analyzed herein are digital scans at 300~dpi, with side lengths
ranging from 1000-2000~pixels.  In this case, each pixel corresponds to 
a length scale on the order of a few 0.1~cm.  Pixels corresponding to a 
target $\Lab$ color (within an allowed color radius) are filtered to form a ``perceived'' representation 
of a particular pattern.
The fractal dimension of the resulting pattern is determined by the
traditional box-counting technique, where
the covering boxes range in size from $d = $1024~px to $d =$4~px,
or length scales of roughly $1.5-2.5~$m to a few millimeters.
The box-counting analysis thus covers about three orders of magnitude.  

The calculated fractal dimensions $D_F$ for both RGB and $\Lab$ spaces 
are displayed in Table~\ref{ciedims}.
What is immediately apparently and interesting
to note is that $\Lab$ space is much more sensitive to changes in
{\it lighter colors}, implying that the calculated dimensions for cream
or white blobs with equal $\beta$ in RGB space will in general {\it not}
be the same in the perceptually-uniform space.  This suggests that the overall 
structure of the blobs may depend on the individual who perceives them, and 
hence the structures may be perceptually different than their physical color
distribution (RGB space) suggests.  Figures~1 and 2
demonstrate how the
physical RGB distribution of a light color is significantly less than the 
perceptual $\Lab$ distribution for the same color.

In fact, for an equal value of $\beta_{\Lab}$, the values of $D_F$ in
$\Lab$ space for lighter colors are consistently higher than the equivalent 
values in RGB space (for fixed $\beta_{\rm RGB}$).  This result in justifiable
based on the nature of the preceptually-uniformity of $\Lab$ space.
In traditional RGB spaces, lighter colors are occupy a much larger volume than
darker colors.  Thus, an analysis which uses a color radius $\beta_{\rm RGB}$ 
will miss significant portions of the space, and will filter a pattern
having a shallower range of ``undistinguishable colors''.  The 
transformation to $\Lab$ space shrinks the volume of the lighter colors
(which correspond to higher luminosity values), thus the associated analysis
will include a much richer depth of colors (and hence a larger pattern will
result)\footnote{An interesting ``test'' of such perceptual distinction
of patterns would be to study the differences in fractal dimensions calculated
from paintings by different artists who largely use subtle, non-luminous
colors.}.

In many cases, the former light color dimensions surpass
the $D_F$ for the darker colors, whereas before they were less than or
equal to them.  If it is true that a viewer will have a preference for
mid-range values of the fractal dimension, $D_F \sim 1.3-1.7$ (as suggested
by the Principle of Aesthetic Middle \cite{fechner} and also supported by recent
data from \cite{taylor3}), then it can be inferred that the darker patterns
``fix'' the fractal dimension for the whole painting.  This is a similar
conclusion to that observed in painting ``construction'' by Taylor {\it et.
al.} \cite{taylor2}, who dubbed this the {\it anchor layer}.

The color spaces used in this analysis correspond to .average. human color 
receptor responses.  Individual variations in these responses, as well as 
those who possess color deficiencies (color-blindness), could certainly 
impact the perceived dimensionality of the patterns.  Indeed, it might be
that the artist himself did not ``see'' the same pattern as did his
audience.  However, color blindness conditions
are more a function of decreased color hue sensitivity, rather than 
luminosity perception (which is the dominant channel in $\Lab$ space).  
Further studies could address these perceptual differences.

As a result, these conclusions can thus be thought of as a preliminary 
assessment of perceptual color fractals.  Further experimentation, 
complemented by psychological behavioral data, is certainly required before
definite conclusions can be made.

\subsection{Choice of Color Scheme and Illuminant}
As previously mentioned, there are numerous possible choices of 
RGB-XYZ transformation matrices used in Equation~(\ref{rgb2xyz}).  These depend on the color
system being used ({\it e.g.} NTSC, PAL), the palette adopted by computer
monitors, and ultimately the standard white defined by the illuminant.
Table~\ref{ciedims} offers a comparison to another D65 illuminant transformation
labeled ``Adobe RGB-XYZ'', having components
\bea
\left( \begin{array}{ccc}
0.576700  & 0.297361  & 0.0270328\\
0.185556  & 0.627355  & 0.0706879\\
0.188212  & 0.0752847 & 0.991248
\end{array} \right) \;
\label{adobergb}
\eea
It is clear from the results that the choice of scheme is mostly inconsequential
to the dimensions being calculated.  Discrepancies can be noted in few
of the color patterns considered.  In fact, these could be explained away
as an improper choice of RGB primaries to begin with.  This cross-comparison
could actually be used as a method for determining the ``actual'' RGB 
coordinates required for the analysis.  In any event, the conclusions from
the previous section are still supported: for a fixed color space radius,
lighter colored patterns will have a perceptually higher fractal dimension
than darker ones. 

\section{Discussion and Conclusions}
Calculating the fractal dimension of patterns based on their RGB coordinates
in the digital representation is not reflective of visual selection criteria
for the same colors due to the non-metric nature of the space.  The $\Lab$
color space is a more natural choice which reflects the color response
of the human perception system, and is a consistent metric space.  This study
has suggested that if the fractal dimensions for dark patterns are in
agreement with previous analysis methods (which they should be, since the
color spaces for darker colors overlap fairly closely), then the lighter
colored patterns possess a much higher fractal dimension approaching $D_F = 2$.
This implies that the distribution of lighter colors -- having higher 
complexity -- would saturate the visual system.  

These results can be 
related to Fechner's ``Principle of the Aesthetic Middle'', which states
that a viewer will tolerate for the longest period of time a visual scene
of moderate complexity \cite{fechner}.  This was experimentally verified by
Berlyne \cite{berlyne1,berlyne2} for statistical distributions, and more
recently applied to fractal analysis by Taylor \cite{taylor3,taylor4}.
The latter reported that human preference for fractals of dimension
$D \sim 1.3$ is the highest.  

However, this work has found that the dimensions for the color patterns 
are significantly above the ``aesthetic middle'' dimension of 1.3.  What
then are the motivations for painting patterns which specifically
are {\it not} aesthetically pleasing to the average viewer?  This is
currently an open question which has no single satisfactory answer.  Borrowing
again from the field of aesthetic research, it is possible to explain Pollock's
choice of dimensions by appealing to the Peak Shift Effect, one of the
``Eight Laws of Artistic Experience'' \cite{rama}.  The Peak Shift Effect
is an experimentally-verified cognitive phenomenon in which visual interest or
identification is strengthened by overtly enhancing key characteristics of
an object or image (such as the ``larger-than-life''
features of caricatures in political cartoons).  These enhanced characterisics
are explicitly {\it not} aesthetically pleasing, but their purpose is to
grab attention and convey key recognition information in a rapid fashion
(see \cite{jrmjsc} for a detailed discussion).

Alternatively,
the relevance to the present work can be understood by considering the relative
difference in fractal dimensions between perceptual colors in Pollock's
work.  That is, based on the notion that lowest fractal dimensions are 
more appealing to observers, this indicates that it is primarily 
the darker patterns play a role in capturing the interest of the observer.
This is consistent with Taylor's earlier notion of the anchor layer,
and in fact serves as a method of ``identifying'' the most salient pattern
on the canvas.  In fact, the ``attractiveness'' of the pattern (based on
lower fractal dimension) and the assertions of this paper could be 
experimentally verified through eye saccade-type or other subject perception
experiments.  

One could speculate that Pollock deliberately ``tuned'' his paintings to contain these color visual structures, based on an intuitive understanding of the visual arts and aesthetics.  This would then indicate a third level of structure in his paintings, in addition to the physical fractals of the paint blobs, as well as the edge fractals created by the luminosity gradients of overlapping pigments \cite{jrmpre}.  If this is indeed true, then it further exemplifies the artistic genius which he demonstrated in creating visually-complex, yet emotionally compelling, non-representational scenes.

\noindent{\bf Acknowledgments}\\
I thank Gerald Cupchik (University of Toronto at Scarborough Division of
Life Sciences) for insightful discussions.

%\pagebreak
%\noindent{\large \bf Figure Captions}\\
%
%\noindent {\bf Figure 1}: Portion of black pigment filter of {\it Autumn Rhythm} showing (a) raw image, (b) physical RGB distribution, and (c) perceptual $\Lab$ distribution \\
%
%\noindent {\bf Figure 2}: Portion of white pigment filter of {\it Autumn Rhythm} showing
%(a) raw image,
%(b) physical RGB distribution, and (c) perceptual $\Lab$ distribution
%corresponding to the data in Table~\ref{ciedims}.
%
\pagebreak

%\begin{table}[h]
%\begin{center}
%{\begin{tabular}{c l c}\hline
%Title (Date) & Dimensions (cm$^2$) \\ \hline
%{\it Blue Poles} (1952) & 486.8 $\times$ 210 \\
%{\it Autumn Rhythm} (1950)& 525.8 $\times$ 266.7 \\
%%{\it Lavender Mist} (1950)& 300 $\times$ 221 \\
%{\it Reflections of the Big Dipper} (1947) & 111 $\times$ 92.1 \\
%{\it Number One A 1948}   & $172.7\times$ 264.2\\
%{\it Number One 1949}  &  160 $\times$ 259 \\ \hline
%\end{tabular}}
%\end{center}
%\caption{
%Catalog of Jackson Pollock images used in analysis \cite{pollock1}.
%}
%\label{paintings}
%\end{table}
%
%\pagebreak
%
\begin{table}[h]
\begin{center}
{\begin{tabular}{|c||c|c|c|}\hline
Color ID & $D_F$ (RGB) & $D_F$ ($\Lab$; sRGB D65) & $D_F$ (Adobe
RGB D65)\\ 
% p32
\hline \multicolumn{4}{|c|}{{\it Reflections of the Big Dipper (1947)}} \\ 
%\multirow{2}{1cm}{I} & 
Black &1.77 & 1.78 (0.04) & 1.77 (0.04) \\
Yellow &1.35 & 1.53 (0.08) & 1.70 (0.06) \\
% p30
%\multirow{2}{1cm}{II} & 
\hline\multicolumn{4}{|c|}{{\it Number One A 1948}} \\ 
Black &1.77 & 1.78 (0.03) & 1.76 (0.04) \\
White &1.57 & 1.79 (0.04) & 1.81 (0.03) \\
% p31
%\multirow{2}{1cm}{III} & 
\hline\multicolumn{4}{|c|}{{\it Undulating Paths}} \\ 
Black &1.76 & 1.75 (0.05) & 1.75 (0.05) \\
Yellow &1.56 & 1.79 (0.04) & 1.80 (0.04) \\
% Lavender Mist, p07
%\multirow{2}{1cm}{IV} & 
\hline\multicolumn{4}{|c|}{{\it Number One 1949}} \\ 
Gray &1.73 & 1.82 (0.03) & 1.83 (0.03) \\
Yellow-gray &1.71 & 1.83 (0.03) & 1.84 (0.03) \\
% Blue Poles
%\multirow{2}{1cm}{V} & 
\hline\multicolumn{4}{|c|}{{\it Blue Poles (1952)}} \\ 
Black &1.74 & 1.49 (0.07) & 1.52 (0.07) \\
Gray &1.68 & 1.78 (0.02) & 1.79 (0.03) \\
% Autumn Rhythm
%\multirow{2}{1cm}{VI} & 
\hline\multicolumn{4}{|c|}{{\it Autumn Rhythm (1950)}} \\ \hline
Black &1.70 & 1.54 (0.05) & 1.51 (0.05) \\
White &1.30 & 1.59 (0.04) & 1.64 (0.03) \\ \hline
\end{tabular}}
\end{center}
\caption{Comparison of fractal dimensions calculated by RGB and $\Lab$
filtering processes for two different RGB-XYZ transformations (D65 illuminants).  
The radii in $\Lab$ color space are chosen to produce
approximately the same value of $D_F$ for darker colors (in this case,
$\beta_{\Lab} = 15$).   The number in parenthesis is the error in the
least-square fit used to calculate the fractal dimension.
}
\label{ciedims}
\end{table}

\begin{figure}[h] \leavevmode
\begin{center}
\includegraphics[scale=0.8]{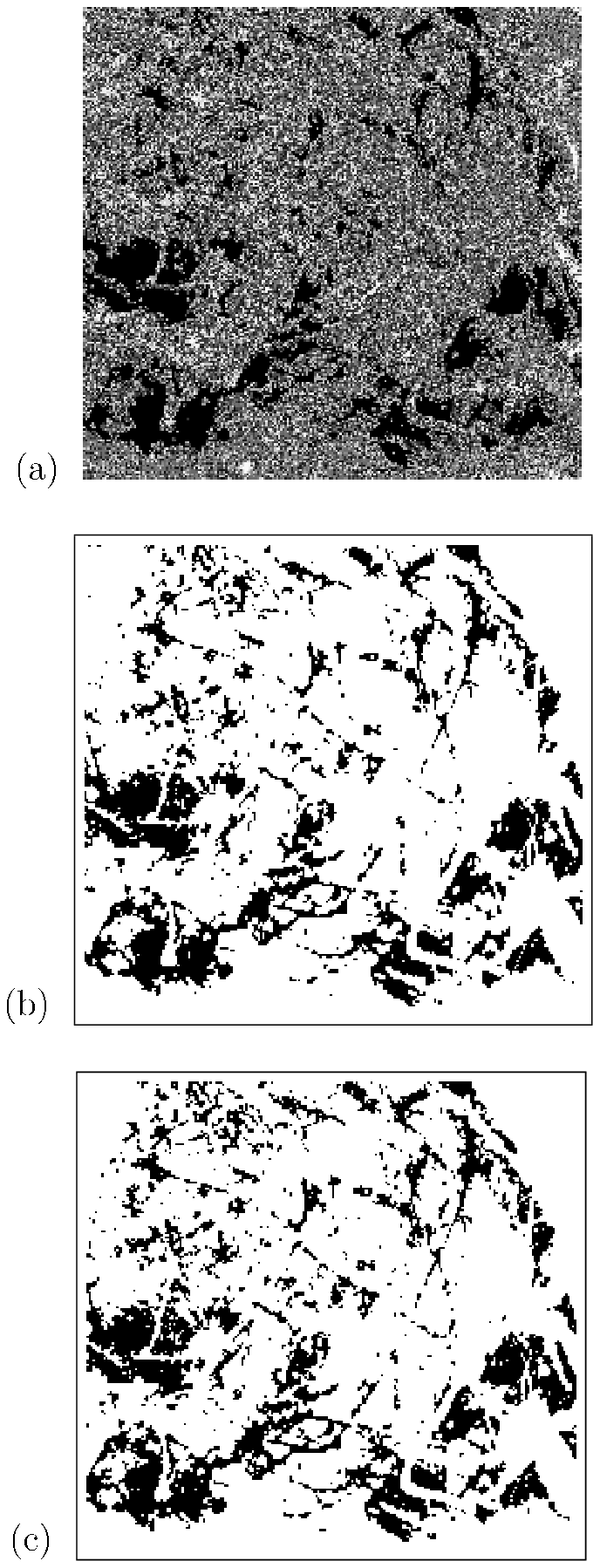}
\caption{Portion of black pigment filter of {\it Autumn Rhythm} showing (a) 
raw image, (b) physical RGB distribution, and (c) perceptual $\Lab$ 
distribution corresponding to the data in Table~\ref{ciedims}.}
\label{fig1}
\end{center}
\end{figure}

\pagebreak
\begin{figure}[h] \leavevmode
\begin{center}
\includegraphics[scale=0.8]{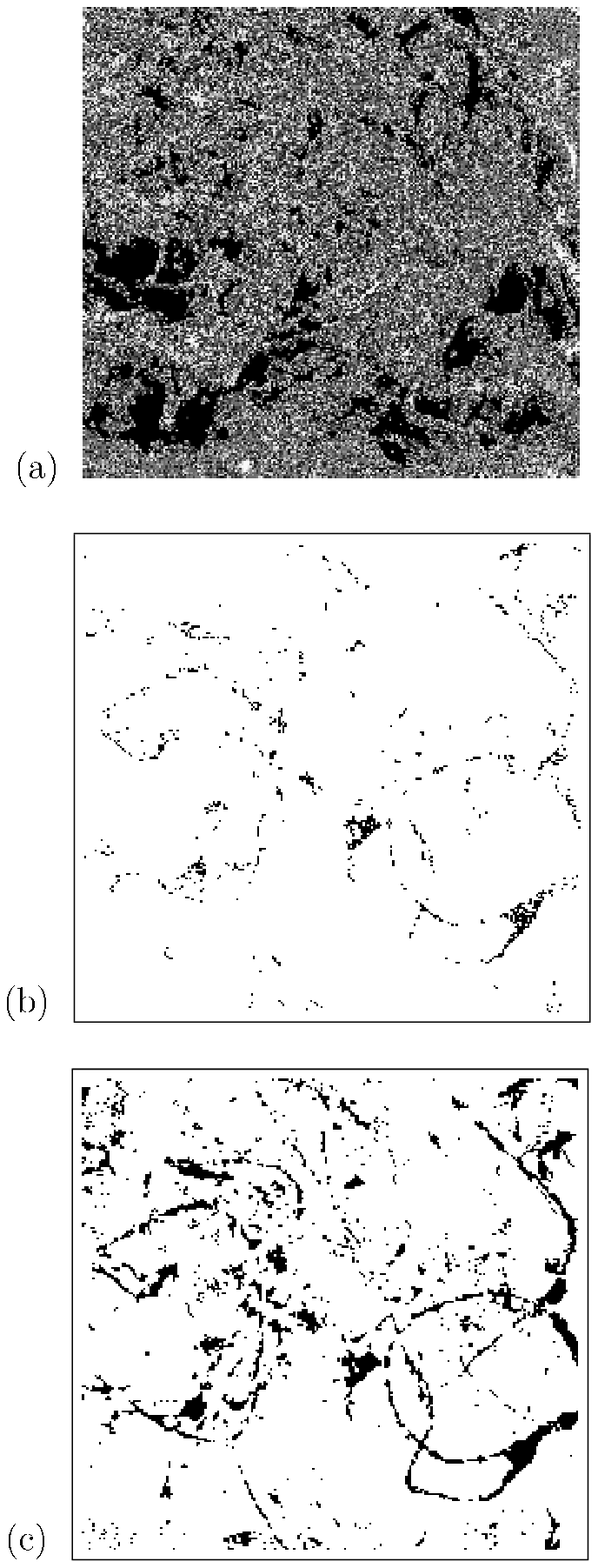}
\caption{Portion of white pigment filter of {\it Autumn Rhythm} showing
(a) raw image,
(b) physical RGB distribution, and (c) perceptual $\Lab$ distribution
corresponding to the data in Table~\ref{ciedims}.}
\label{fig2}
\end{center}
\end{figure}

\end{document}